\documentclass[%
aps,
pre,
reprint,
superscriptaddress,
twocolumn,
showpacs,
amsmath,
amssymb,
amsfonts,
a4paper,
]{revtex4}

\usepackage{subfigure}
\usepackage{graphicx}
\usepackage{dcolumn}
\usepackage{bm}
\usepackage{latexsym}
\usepackage{color}

\newcommand{\be}{\begin{equation}}
\newcommand{\ee}{\end{equation}}
\newcommand{\R}{{\mathbb{R}}}

\begin{document}

\title{Reproduction time statistics and segregation patterns in growing populations}

\author{Adnan Ali}
\affiliation{%
Centre for Complexity Science, University of Warwick,
Coventry CV4 7AL, United Kingdom
}
\author{Ell\'ak Somfai}
\affiliation{%
Centre for Complexity Science, University of Warwick,
Coventry CV4 7AL, United Kingdom
}
\affiliation{%
Department of Physics, University of Warwick,
Coventry CV4 7AL, United Kingdom
}
\author{Stefan Grosskinsky}
\affiliation{%
Centre for Complexity Science, University of Warwick,
Coventry CV4 7AL, United Kingdom
}
\affiliation{%
Mathematics Institute, University of Warwick,
Coventry CV4 7AL, United Kingdom
}

\date{\today}

\begin{abstract}
Pattern formation in microbial colonies of competing strains under purely space-limited population growth has recently attracted considerable research interest. We show that the reproduction time statistics of individuals has a significant impact on the sectoring patterns. Generalizing the standard Eden growth model, we introduce a simple one-parameter family of reproduction time distributions indexed by the variation coefficient $\delta\in [0,1]$, which includes deterministic ($\delta =0$) and memory-less exponential distribution ($\delta =1$) as extreme cases. We present convincing numerical evidence and heuristic arguments that the generalized model is still in the KPZ universality class, and the changes in patterns are due to changing prefactors in the scaling relations, which we are able to predict quantitatively. At the example of \textit{Saccharomyces cerevisiae}, we show that our approach using the variation coefficient also works for more realistic reproduction time distributions.
\end{abstract}

\maketitle

\section{Introduction}

Spatial competition is a common phenomenon in growth processes and can lead to
interesting collective phenomena such as fractal geometries and pattern
formation \cite{Lacasta1999,Matsushita1999,Barabasi1995}. This is relevant in
biological contexts such as range expansions of biological species
\cite{Wegmann2006,Fisher2001} or growth of cells or microorganisms, as well as
in social contexts such as the dynamics of human settlements or urbanization
\cite{Finlayson2005}. These phenomena often exhibit universal features which do
not depend on the details of the particular application, and have been studied
extensively in the physics literature
\cite{Matsushita1999,Vicsek1990,Avnir1989,BenJacob1990,Meakin1998,Barabasi1995}.

Our main motivating example will be growth of microbes in two dimensional
geometries, for which recently there have been several quantitative studies. In
general, the growth patterns in this area are influenced by many factors, such
as size, shape and motility of the individual organism \cite{Wakita1997}, as
well as environmental conditions such as distribution of resources and
geometric constraints \cite{Ohgiwari1992,Thompson1992}, which in turn influence
the proliferation rate or motility of organisms \cite{Tokita2009}. We will
focus on cases where active motion of the individuals can be neglected on the timescale of growth, which
leads to static patterns and is also a relevant regime for range
expansions. We further assume that there is no shortage of resources, and growth
and competition of species is purely space limited and spatially homogeneous.
This situation can be studied for colonies of immotile microbial species grown under precisely
controlled conditions on petri dish with hard agar and rich growth medium.

Under these conditions one expects the colony to form compact Eden-type clusters
\cite{Ohgiwari1992}, which has recently been shown for various species including
\textit{Saccharomyces cerevisiae}, \textit{Escherichia coli}, \textit{Bacillus subtilis} and \textit{Serratia marcescens} 
\cite{Hallatschek2007,Tokita2009}.

\begin{figure}[t]
\begin{center}
\subfigure[]{\label{fluorescent1}\includegraphics[height=1.15in,width=1.5in]{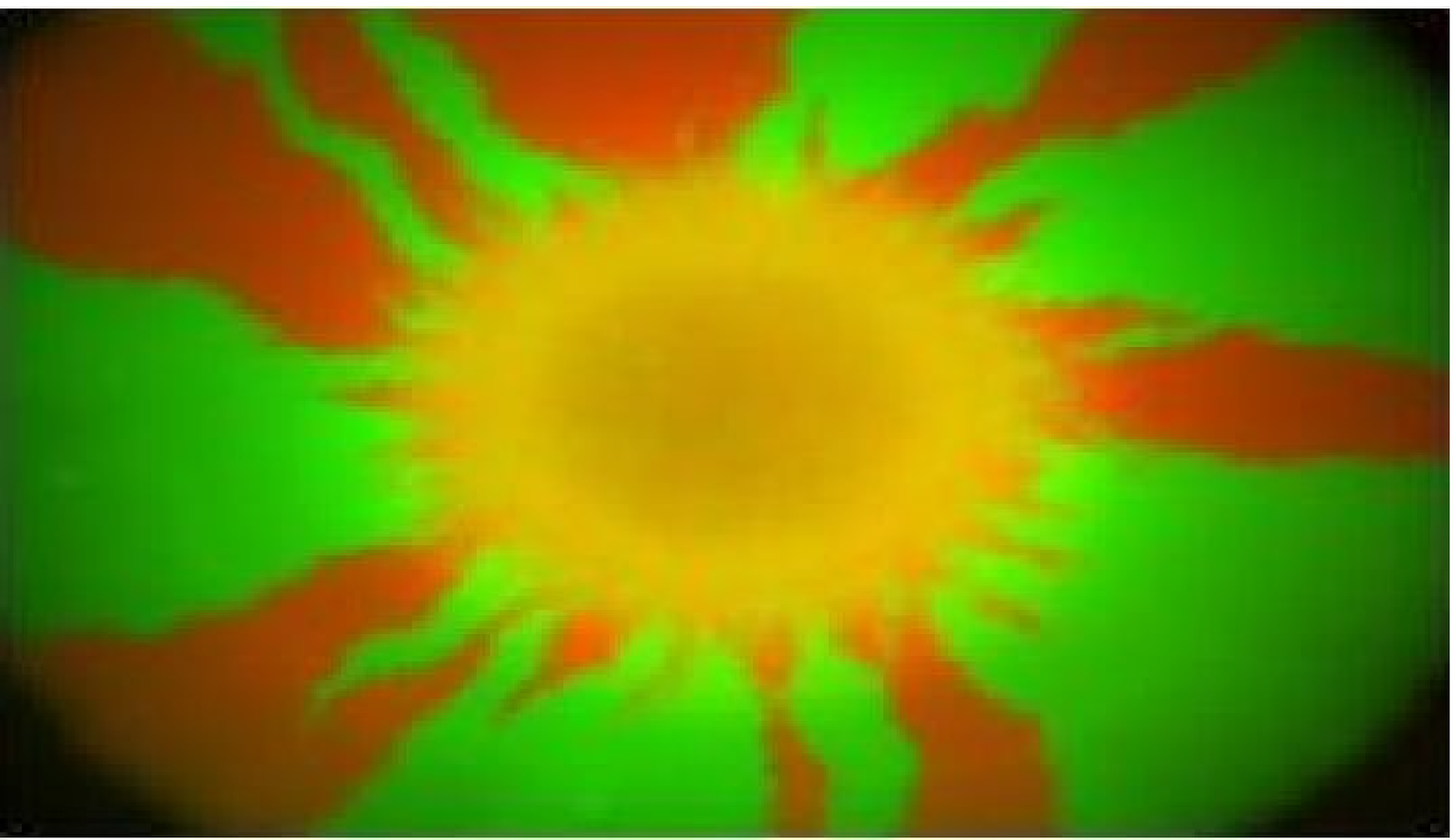}}
\subfigure[]{\label{fluorescent2}\includegraphics[width=1.5in]{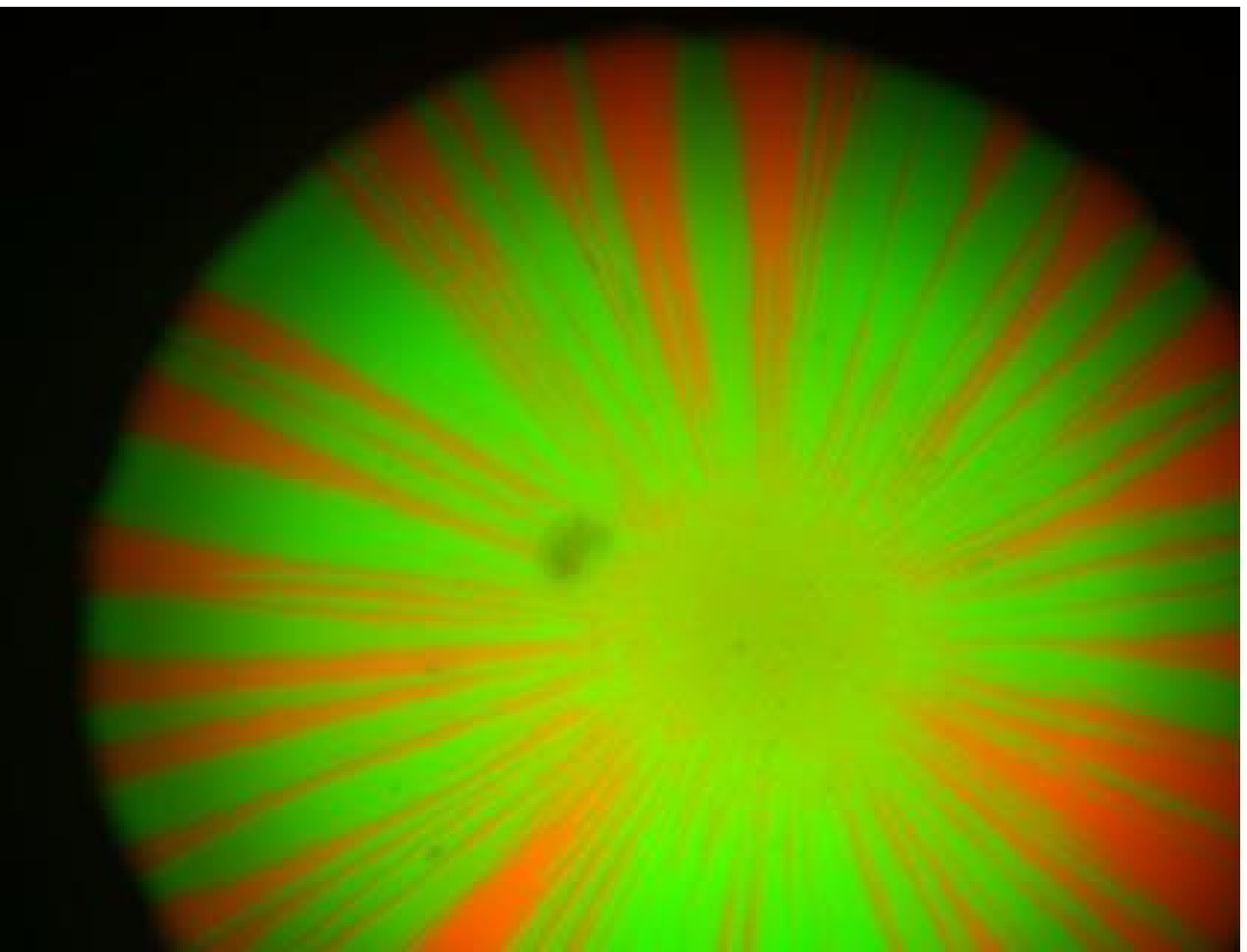}}
\end{center}
\caption{\label{fig:fluorescent}
(Color online) Fluorescent images of colonies of (a) \textit{E.~coli} and (b) \textit{S.~cerevisiae}. The scaling properties of both patterns are believed to be in the KPZ universality class, and the differences are due to microscopic
details of the mode of reproduction and shape of the micro-organisms. The images have been taken with permission from 
\cite{Hallatschek2007}, copyrighted (2007) by the National Academy of Sciences, U.S.A. 
}
\end{figure}

The Eden model \cite{Eden1961} has been introduced as a basic model for the
growth of cell colonies. It has later been shown to be in the KPZ universality
class \cite{Vicsek1991,Barabasi1995,Vicsek1990,Kardar1986}, which describes the scaling
properties of a large generic class of growth models. In recent detailed
studies of \textit{E.~coli} and \textit{S.~cerevisiae}
\cite{Hallatschek2007,Tokita2009,Grosskinsky2010,Hallatschek2010} quantitative evidence for the
KPZ scaling of growth patterns has been identified. The models used in these
studies ignored all microscopic details reproduction, such as anisotropy of cells \cite{Slaughter2009}, and could
therefore not explain or predict differences observed for different species. 
Nevertheless, they provided a good reproduction of the basic features such as KPZ
exponents, which is a clear indication that segregation itself is an emergent phenomenon.
Fig.~\ref{fig:fluorescent} shows differences in growth patterns in a circular
geometry taken from \cite{Hallatschek2007} for immotile \textit{E.~coli} and \textit{S.~cerevisiae}.  
For both species the microbial populations are made of two strains, which
are identical except having different fluorescent labeling. Reproduction is
asexual, and the fluorescent label carries over to the offspring. At the beginning of the experiments the strains are well mixed, but during growth rough sector shaped segregated regions develop. The qualitative emergence of these segregation patterns and connections to annihilating diffusions has been studied in \cite{Grosskinsky2010,Hallatschek2007,Hallatschek2010,Korolev2010}, ignoring all details specific for a particular species.

For \textit{S.~cerevisiae} the domain boundaries are less rough when compared to
\textit{E.~coli}, leading to a finer pattern consisting of a larger number of sectors. In
general, this is a consequence of differences in the mode of reproduction and
shapes of the microbes, which introduce local correlations that are not present
in simplified models. In this paper we focus on the effect of time correlations introduced by
reproduction times that are not exponentially distributed (as would be in
continuous time Markovian simulations), but have a unimodal distribution with
smaller variation coefficient. This is very relevant in most biological
applications (see e.g. \cite{Cole2004,Cole2007,Karen2008}), and even in spatially isotropic systems the resulting temporal correlations lead to more regular growth and
therefore smaller fluctuations of the boundaries, with an effect on the patterns as seen in Fig.~\ref{fig:circular}.

To systematically study these temporal correlations, we introduce a generic 
one-parameter family of reproduction times, explained in
detail in Section \ref{sec:model}. We establish that the growth clusters and
competition interfaces still show the characteristic scaling within the KPZ
universality class, and the effect of the variation coefficient is present only
in prefactors. We predict these effects quantitatively and find good agreement
with simulation data; these results are presented in Section \ref{sec:results}.
More realistic reproduction time distributions with a higher number of
parameters are considered in Section \ref{sec:bio}, where we show that to a
good approximation the effects can be summarized in the variation coefficient
and mapped quantitatively onto our generic one-parameter family of reproduction
times. Therefore, our results are expected to hold quite generally for unimodal
reproduction time distributions, and the variation coefficient alone determines 
the leading order statistics of competition patterns.

\section{\label{sec:model}The Model}

For regular reproduction times with small variation coefficient the use of a
regular lattice would lead to strong lattice effects that affect the shape of
the growing cluster. To avoid these we use a more realistic Eden growth model in a continuous
domain in $\R^2$ with individuals modelled as circular hard-core particles with
diameter $1$, since we want to study purely the effect of time correlations.
This leads to generalized Eden clusters which are compact with an interface
that is rough due to the stochastic growth dynamics.

\begin{figure}[t]
\begin{center}
\subfigure[]{\label{circular1}\includegraphics[]{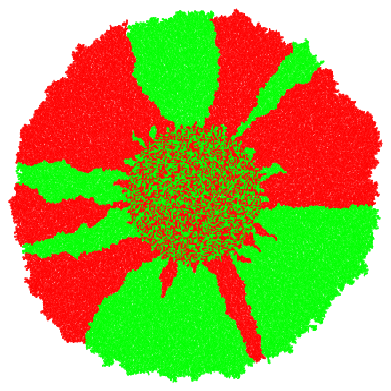}}
\subfigure[]{\label{circular2}\includegraphics[]{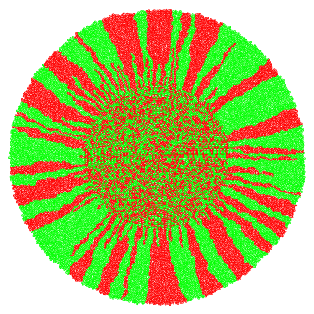}}
\end{center}
\caption{\label{fig:circular}
(Color online) A smaller variation coefficient $\delta$ in reproduction times (see (\ref{repro}) and (\ref{delta})) leads to more regular growth, smoother domain boundaries and finer sectors. Shown are simulated circular populations with (a) $\delta=1$ and (b) $\delta=0.1$. Both colonies have an initial radius of
$r_{0}=50$, and they are grown up to simulation time $t=50$ leading to final radii of approximately $120$ (a) and $95$ (b). The different colors denote cell types $1$ and $2$.
}
\end{figure}

Let $B(t)$ denote the general index set of particles $p$ at time $t$,
$(x_p ,y_p )\in\R^2$ is the position of the centre of particle $p$,
and $s_p \in\{ 1,2\}$ is its type. We write $B(t)=B_1 (t)\cup B_2 (t)$ as the union of the sets of
particles of type $1$ and $2$.  We also associate with each particle
the time it tries to reproduce next, $T_p >0$.  Initially, $T_p$ are i.i.d. random variables with cumulative distribution $F_\delta$ with parameter $\delta\in [0,1]$, which is
explained in detail below.  After each reproduction $T_p$ is incremented by a
new waiting time drawn from the same distribution.
Note that we focus entirely on the neutral case, i.e. the reproduction time is
independent of the type and both types have the same fitness.  We describe
the dynamics below in a recursive way.

Following a successful reproduction event of particle $p$ at time $t=T_p$, a new
particle with index $q=|B(T_p-)|+1$ is added to the set $B_{s_q}$ with the same cell type $s_q =s_p$, such that
\be
B_{s_p} (T_p+)=B_{s_p} (T_p-) \cup \{ q\}\ .
\ee
Here $B(T_p-)$ and $B(T_p+)$ denote the index set just before and just after the reproduction event, and $|B(t)|$ denotes the size of the set $B(t)$.
The position of the new particle is given by
\be
(x_q ,y_q)=(x_p ,y_p )+(\cos\phi ,\sin\phi)\ ,
\ee
where $\phi\in [0,2\pi )$ is drawn uniformly at random. This is subject to a
hard-core exclusion condition for circular particles, i.e. the Euclidean
distance to all other particle centres has to be at least $1$, as well as to other constraints depending on the simulated geometry as explained below. Note that in our model the daughter
cell touches its mother, which is often realistic but in fact not essential,
and the distance could also vary stochastically over a small range. The
new reproduction times of mother and daughter are set as
\be
T_p \mapsto T_p^\text{old} +T\ ,\quad T_q =T_p^\text{old} +T'\ ,
\ee
where $T,T'$ are i.i.d. reproduction time intervals with distribution
$F_\delta$. There can be variations on this where mother and daughter have
different reproduction times, which are discussed in Section \ref{sec:bio}. The
next reproduction event will then be attempted at $t=\min\big\{ T_q :q\in
B(T_p+)\big\}$. Reproduction attempts can be unsuccessful, if there is no
available space for the offspring due to blockage by other particles.  In this
case the attempt is abandoned and $T_p$ is set to $\infty$, as due to the
immotile nature of the cells this particle will never be able to reproduce.

The initial conditions for spatial coordinates and types depend on the
situation that is modelled. In this paper we mostly focus on an upward growth
in a strip of length $L$ with periodic boundary conditions on the sides, where
we take $B(0)=\{ 1,\ldots ,L\}$ with $(x_p, y_p) =(p,0)$,  for all $p\in B(0)$.
The initial distribution of types can be either regular or random depending on
whether we study single or interacting boundaries, and will be specified later.

In Section \ref{sec:results} for our main results we use reproduction times $T$ distributed as
\be\label{repro}
1-\delta +\text{Exp}(1/\delta )\ ,\quad\delta\in (0,1]\ ,
\ee
i.e. $T$ has an exponential distribution with a time lag $1-\delta\in [0,1)$ and a mean fixed to $\langle T\rangle =1$ for all $\delta$. The corresponding cumulative distribution function $F_\delta$ is given by
\be
F_\delta (t)=\left\{\begin{array}{cl} 0 &,\ t\leq 1-\delta\\ 1-e^{-(t-1+\delta)/\delta} &,\ t\geq 1-\delta
\end{array}\right.\ .
\ee
The variation coefficient of this distribution is given by the standard deviation divided by the mean, which turns out to be just
\be\label{delta}
\frac{\sqrt{\langle T^2\rangle -\langle T\rangle^2}}{\langle T\rangle}=\frac{\delta}{1}=\delta\ .
\ee
With this family we can therefore study reproduction which is more regular then
exponential with a fixed average growth rate of unity (equivalent
of setting the unit of time).

\begin{figure}[t]
\begin{center}
\subfigure[]{\label{linear1}\includegraphics[bb=103 87 359 272,clip,width=1.5in,height=1.2in]{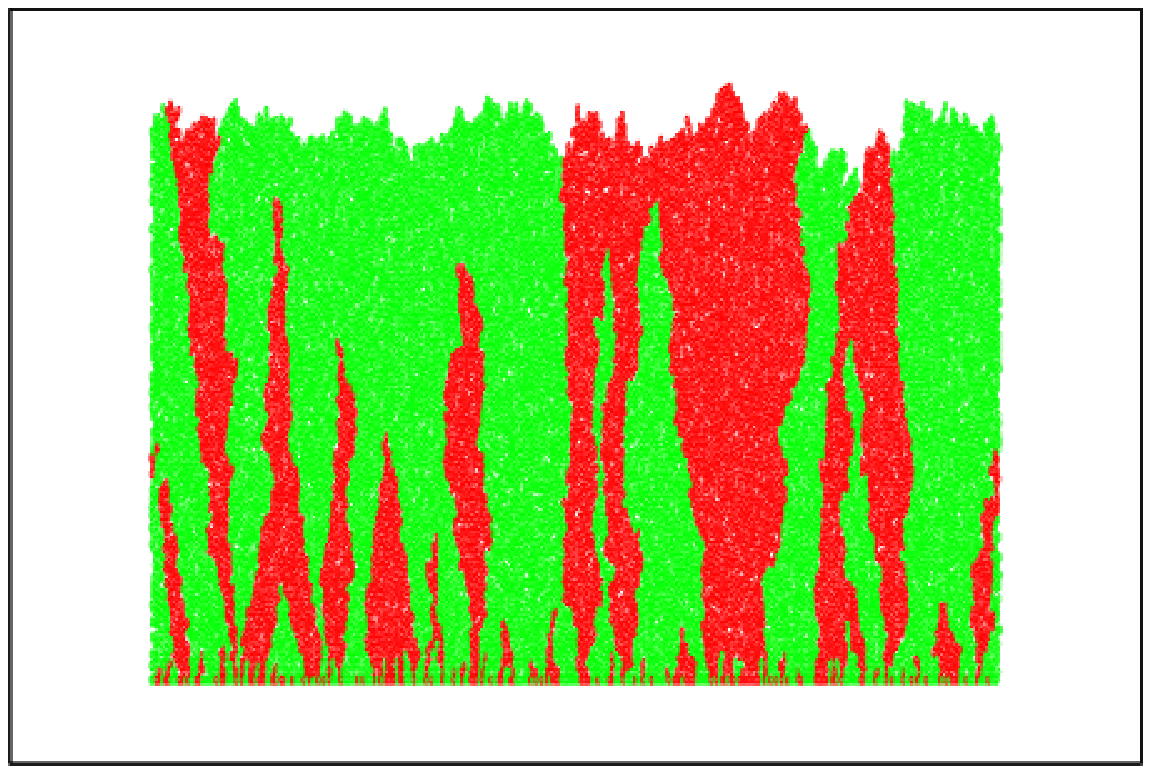}}
\subfigure[]{\label{linear2}\includegraphics[bb=103 98 359 274,clip,width=1.5in,height=0.7in]{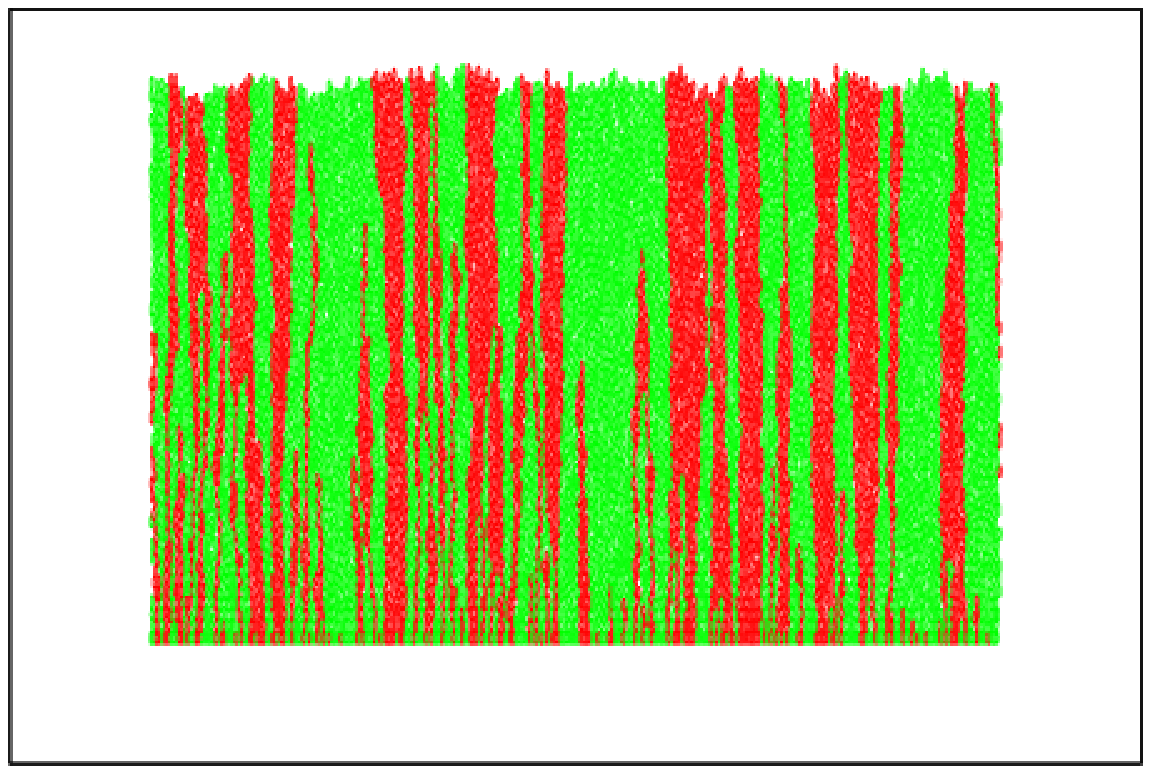}}
\end{center}
\caption{\label{fig:linear}
(Color online) Populations in a linear geometry with periodic boundary conditions in lateral direction with (a) $\delta=1$ and (b) $\delta=0.1$. Both populations have lateral width $L=300$, and the colonies are grown to a simulation time $t \approx 50$, leading to heights of approximately $70$ (a) and $40$ (b). The different colors denote cell types $1$ and $2$.}
\end{figure}

For $\delta =1$ this is a standard Eden cluster, but $\delta <1$ introduces
time correlations.  While the correlations affect the fluctuations, we present
convincing evidence that they decay fast enough not to change the
fluctuation exponents, so the system remains in the KPZ universality class.
Furthermore we make quantitative predictions on the $\delta$-dependence of
non-universal parameters and compare them to simulations. The more
synchronized growth leads to effects similar to the ones seen in experiments
(Fig.~\ref{fig:fluorescent}). To give a visual impression of the patterns
produced by the model, we show in Fig.~\ref{fig:circular} two growth
patterns with $\delta =1$ and $0.1$. The initial condition is a circle, and
the types are distributed
uniformly at random. The patterns are qualitatively similar to the experimental
ones in Fig.~\ref{fig:fluorescent}, and more regular growth leads to a finer sector
structure. The same effect is shown on Fig.~\ref{fig:linear} for the
simulations in a linear geometry with periodic boundary conditions, which is
analyzed quantitatively in the next Section. Smaller values of $\delta$ also lead to more compact growth and smaller height values reached in the same time.

\section{\label{sec:results}Main Results}

\subsection{Quantitative analysis of the colony surface }

In this Section we provide a detailed quantitative analysis of the $\delta$ family
of models in linear geometry with periodic boundary conditions 
(see Fig.~\ref{fig:linear}), starting with the dynamical scaling properties of the growth interface.

We regularize the surface to be able to define it as a function of the lateral coordinate $x$ and time $t$ as
\be
y(x,t):=\max\big\{ y_p :p\in B(t), |x_p -x|\leq 1\big\}\ .
\ee
In case of overhangs (which are very rare) we take the largest possible height, and due to the discrete nature of our model this leads to a piecewise constant
function.

The surface of a standard Eden growth cluster is known to be in the KPZ
universality class \cite{Kardar1986,Eden1961}, i.e. a suitable scaling limit of $y(x,t)$ with vanishing particle diameter fulfills the KPZ equation
\be
  \partial_{t} y(x,t)= v_0+\nu \Delta y(x,t)
  + \frac{\lambda}{2}(\nabla y(x,t))^{2} + \sqrt{D}\eta(x,t)
  \label{kpz} .
\ee
Here $v_0$, of the order of unity, corresponds to the growth rate of the
initial flat surface (related to the mean reproduction rate and some
geometrical effects), the surface tension term with $\nu >0$ represents surface
relaxation, and the nonlinear term represents the lowest order contribution to lateral growth \cite{Kardar1986}. The fluctuations due to stochastic growth are
described by space-time white noise $\eta(x,t)$, which is a mean $0$ Gaussian
process with correlations
\be
    \big\langle \eta(x,t)\eta(x',t') \big\rangle =\delta(x-x')\delta(t-t').
\ee
We denote the average surface height by
\be
    h(t) :=\frac1L\int_0^L y(x,t)\, dx\ ,
\ee
which is a monotone increasing function in $t$. It is also
asymptotically linear and therefore we will later also use $h$ as a proxy for time. The $\delta$-dependence of the average growth velocity of height as seen in Fig.~\ref{fig:linear} does not lead to leading order contributions to the statistical properties of the surface or the structure of sectoring patterns.

\begin{figure}[t]
\begin{center}
\includegraphics[clip,width=0.9\columnwidth]{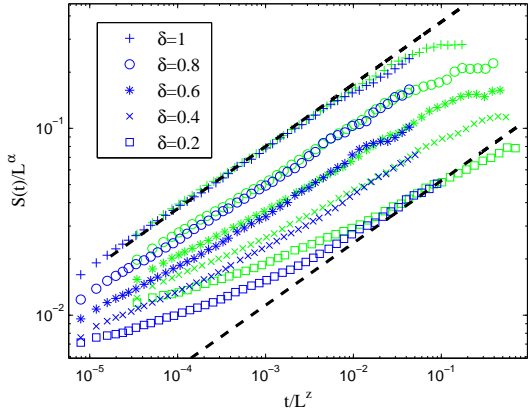}
\end{center}
\caption{\label{roughness} 
    (Color online) Family-Vicsek scaling (\ref{Family}) of the surface roughness $S(t)$. The
    data collapse under rescaling with $\alpha =1/2$ and $z=3/2$ occurs in a scaling window which is narrower for small $\delta$ due to intrinsic correlations. The different
    symbols correspond to different values of $\delta$, and the color
    represents system size, $L=1500$ (green/light grey) and $L=4000$ (blue/dark grey). The
    dashed lines indicate the expected slope $\beta =1/3$. The data for $L=1500$ has been averaged over $100$ samples and for $L=4000$ over $30$ samples. The error bars are comparable to the size of the symbols. 
}
\end{figure} 

The roughness of the surface is given by the root mean
squared displacement of the surface height as a function of $t$
\cite{Meakin1998,Barabasi1995}, defined as 
\be
    S(t)=\Big\langle\frac1L\int_0^L \big( y(x,t)-h(t)\big)^2 dx\Big\rangle^{1/2}
    \ .
\ee
The main properties of the surface $y(x,t)$ can then be characterized by the
Family-Vicsek scaling relation of the roughness
\begin{equation}
    S(t)=L^{\alpha} f(t/L^z) \label{Family}\ ,
\end{equation}
where the scaling function $f(u)$ has the property
\begin{equation}
    f(u) \propto \left \{ \begin{array}{lr} u^{\beta} & \hspace{1cm} u\ll 1 \\
     1 & \hspace{1cm} u\gg1\end{array} \right.\label{scalingfunction}.
\end{equation}
Such a scaling behaviour has been shown for many discrete models including
ballistic deposition and continuum growth \cite{Huergo2010,Meakin1998,Barabasi1995,Kardar1986,Family1985}, and holds also for other universality classes such as Edward Wilkinson. For
the KPZ class in $1+1$ dimensions the saturated interface roughness exponent
is $\alpha=1/2$, the growth exponent is $\beta=1/3$, and the dynamic
exponent is $z=\alpha /\beta =3/2$.

Fig.~\ref{roughness} shows a data collapse for the roughness $S(t)$ for two
system sizes, and for a number of different values of $\delta$.  As
$\delta$ gets smaller, the surface becomes less rough due to a more
synchronized growth. The dashed lines indicate the power law growth with
exponent $\beta =1/3$ in the scaling window. This window ends around $t/
L^z\approx 1$ due to finite size effects, where the lateral correlation length
reaches the system size and the surface fluctuations saturate.  For small
$t$ the system exhibits a transient behaviour before entering the KPZ scaling
due to local correlations resulting from the non-zero particle size and stochastic growth rules. As we quantify later, these
correlations are much higher for more synchronized growth at small $\delta$,
which leads to a significant increase in the transient regime. The transient time scale is independent of system size and vanishes in the scaling limit, so that  
the length of the KPZ scaling window increases with $L$. This behaviour can be observed in Fig.~\ref{roughness} where for the smallest value $\delta =0.2$ the scaling regime is still hard to identify for the accessible system sizes.

Another characteristic quantity is the height-height correlation function
defined as \cite{Barabasi1995,Krug1992,Amar1992}
\begin{equation}
    C(l,t)=\Big\langle\frac1L
    \int_0^L (y(x,t)-y(x+l,t))^{2} dx\Big\rangle^{1/2} 
    \label{correlationfunction}.
\end{equation}

\begin{figure}[t]
\begin{center}
\includegraphics[width=0.9\columnwidth]{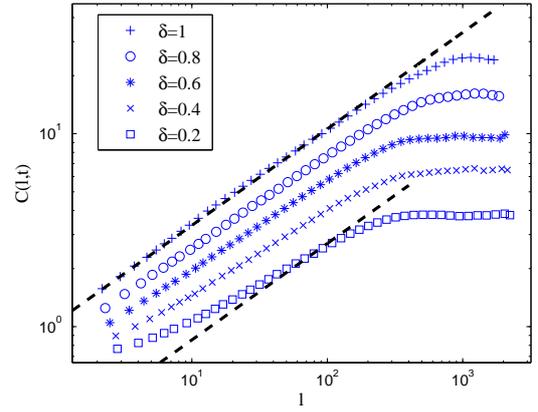}
\end{center}
\caption{\label{correlation}
 (Color online) The height-height correlation function $C(l,t)$ for $L=4000$ at
    $t=11000$ for various values of $\delta$. The data has been averaged over $30$ samples, and the error bars are comparable to the size of the symbols. The
    dashed lines indicate the expected slope $1/2$.
    }
\end{figure}
For a KPZ surface in $1+1$ dimensions this obeys the scaling behaviour
\begin{equation}
    C(l,t) \sim \left \{ \begin{array}{lr} 
        (\frac{D}{2\nu} l)^{1/2} & \hspace{1cm} l \ll \xi_{\parallel}(t) \\
        \left(\frac{D}{2\nu}\right)^{2/3}  (\lambda t)^{1/3} &
            \hspace{1cm}  l \gg \xi_{\parallel}(t) \end{array}\right.
    \label{correlationresult},
\end{equation}
where $\xi_{\parallel}(t)$ is defined to be the lateral correlation length
scale and takes the form \cite{Barabasi1995,Rost1995,Amar1992}
\be\label{paral}
    \xi_{\parallel}(t)\sim (D/2\nu)^{1/3}(\lambda t)^{2/3} \ .
\ee
A detailed computation can be found in Appendix \ref{derivecorrelation}. 
For small values $C(l,t)$ grows as a power-law with $l$, and when $l$ exceeds the lateral correlation length $\xi_{\parallel}(t)$ it reaches a value that depends on the time
$t$ and the parameters of (\ref{kpz}). This is shown in Fig.~\ref{correlation}, where
$C(l,t)$ is plotted for various values of $\delta$,  
and the data agree well with the exponent $\alpha =1/2$ for the KPZ class indicated by dashed lines.

\begin{figure}[t!]
\begin{center}
\includegraphics[width=0.9\columnwidth]{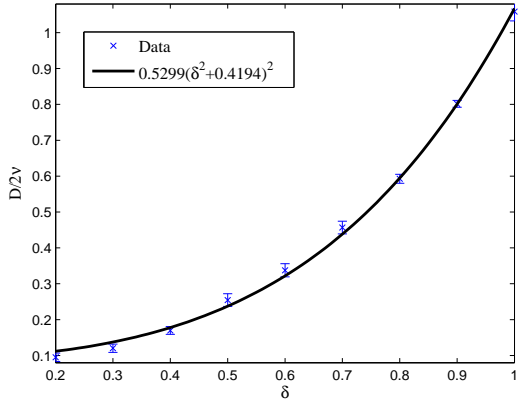}

\end{center}
\caption{\label{KPZcoefficient} 
    (Color online) Dependence of the KPZ parameters $D/(2\nu)$ on $\delta$. Data are obtained from (\ref{correlationresult}) by fitting the prefactor of the power law in Fig.~\ref{correlation}, using that the proportionality constant is very close to $1$ (cf. derivation (\ref{exactpre}) in the Appendix). The data are in good agreement with the prediction (\ref{d2nu}) with fitted parameters $\epsilon \approx 0.42$ and $D/(2\nu)(\delta =1) \approx 1.1$.
}
\end{figure}

\newcommand{\var}{{\operatorname{var}}}

The time correlations introduced by the partial synchronization 
can be estimated by considering a chain of $N$ growth
events where each particle is the direct descendant of the previous one. Each 
added particle corresponds to a height change $\Delta y_i$, and has an
associated waiting time $T_i$ with distribution (\ref{repro}). During time $t$ there are $N(t)$ growth events, and since the average reproduction time is $1$ with variance $\delta^2$, we have $\langle N(t)\rangle \approx t$ and $\var (N(t)) \approx \delta^2 t$. The height of the last particle is $y_{N(t)}=\sum_i^{N(t)}\Delta y_i$, leading to
\be
\var(y_{N(t)}) =\langle\Delta y_i \rangle^2 \,\var(N(t))+\langle N(t)\rangle\, \var(\Delta y_i )\ .
\ee
The terms in this expression correspond to two sources of uncertainty: (i) due to the randomness in $T_i$ the number of growth events vary with $\var(N(t))$, and (ii) the individual height increments are random with $\var(\Delta y_i)$. This leads to
\be
\var(y_{N(t)}) \approx t\,\langle\Delta y_i \rangle^2 (\delta^2 + \epsilon^2 )\ ,
\ee
where $\epsilon =\sqrt{\var(\Delta y_i )}/\langle\Delta y_i \rangle$ denotes the variation coefficient of the height fluctuations due to geometric effects.

We define the correlation time $\tau$ as the amount of time by which 
the uncertainty of the height of the chain becomes comparable to one
particle diameter, $\var(y_{N(\tau)})=O(1)$. Since $\langle \Delta y_i \rangle$ is largely independent of $\delta$ (cf.\ Appendix \ref{epsilon}), the time correlation induces a fixed intrinsic vertical correlation length
\be\label{xiperp}
    \tau \sim \frac{1}{\delta^2 + \epsilon^2}
\ee
in the model. This correlation length reduces fluctuations and leads to an increase in the saturation time $t_{sat}$ of the system, namely $t_{sat} /\tau \sim L^z$, a modification of the usual relation with the system size $L$. Analogous to the standard derivation of the time-dependence of the lateral correlation length \cite{Barabasi1995}, this leads to
\be
\xi_{\parallel}(t)\sim (t/\tau)^{1/z}\ .
\ee
Together with (\ref{paral}) from the behaviour of the correlation length, we expect
\be\label{d2nu}
D/(2\nu )\sim (\delta^2 +\epsilon^2 )^2 \ ,
\ee
since $\lambda$ turns out to be largely independent of $\delta$. 
This is shown to be in very good agreement with the data in
Fig.~\ref{KPZcoefficient}, for fitted values of $\epsilon$ and a prefactor. The fit value for $\epsilon$ and the ratio $D/(2\nu )$ for $\delta =1$ (the usual Eden model) are compatible with simple theoretical arguments (see Appendix \ref{epsilon}). So the very basic argument above to derive an intrinsic vertical correlation length explains the $\delta$-dependence of the surface properties very well. Measuring height in this intrinsic length scale, we observe a standard KPZ behaviour with critical exponents being unchanged, since the intrinsic correlations are short range (i.e. decay exponentially on the scale $\tau$). This is in contrast to effects of long-range correlations where the exponents typically change, see e.g. studies with long-range temporally correlated noise \cite{Medina1989,Amar1991,Katzav2004} or memory and delay effects using fractional time derivatives and integral/delay equations \cite{Xia2011,Gang2010,Chattopadhyay2009}.

\begin{figure}[ht!]
\begin{center}
\subfigure[]{\label{boundary1}\includegraphics[clip,width=0.9\columnwidth]{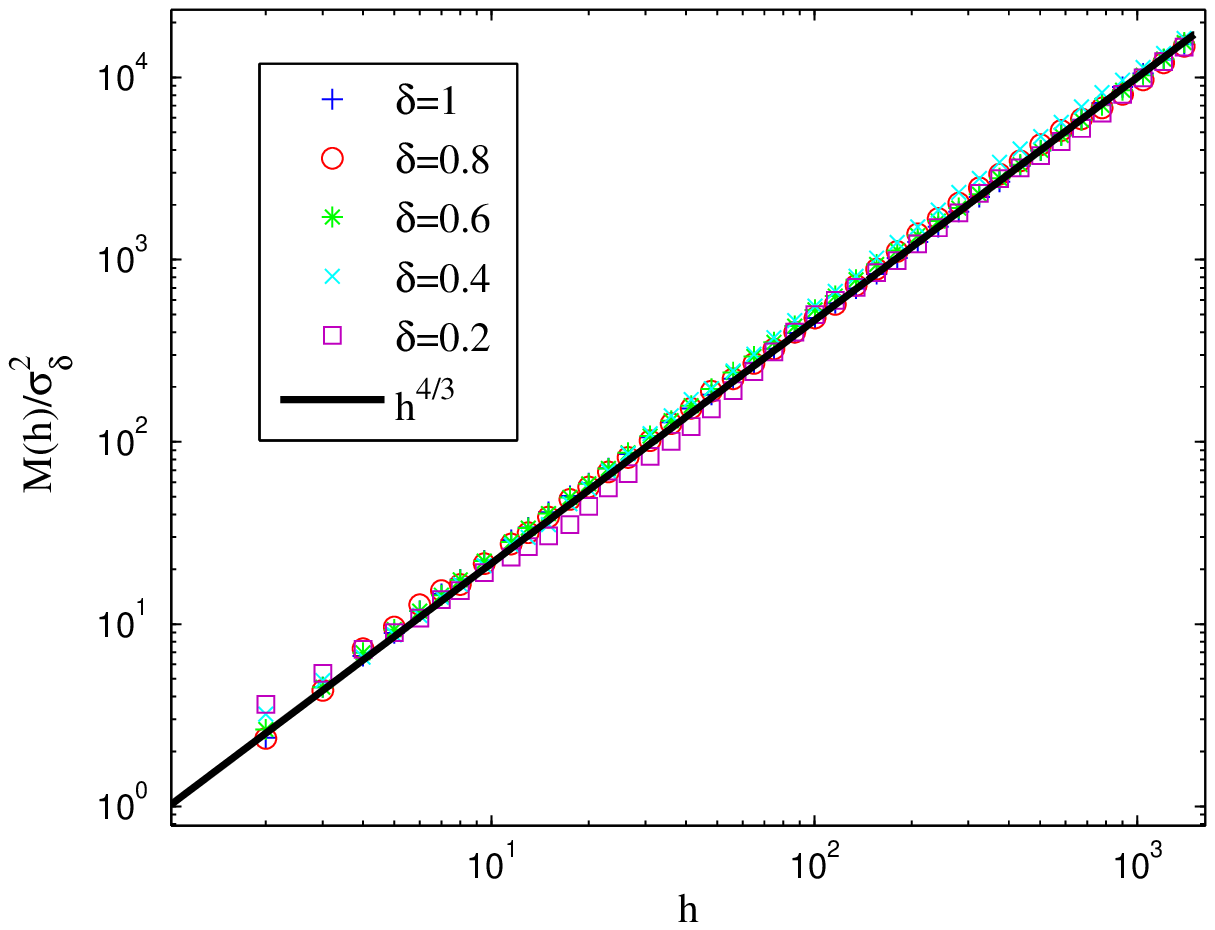}}
\subfigure[]{\label{boundary2}\includegraphics[clip,width=0.9\columnwidth]{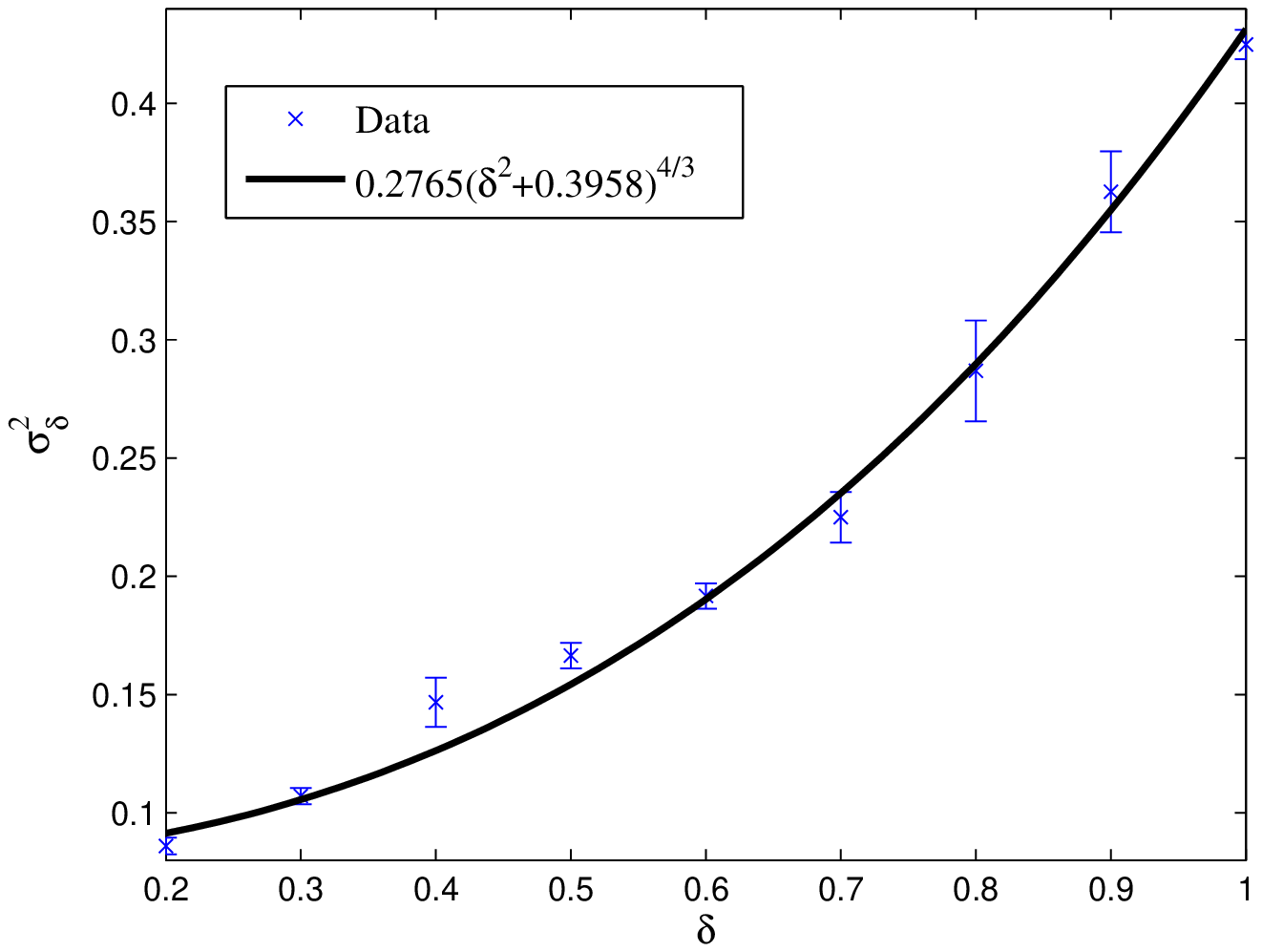}}
\end{center}
\caption{\label{boundary}
(Color online) Scaling behaviour of the mean square displacement $M(h)$ (\ref{msd2}). The system size is $L=1000$ and the data is averaged over $500$ samples and the error bars are comparable to the size of the symbols. 
(a) Data collapse of the normalized quantity $M(h)/\sigma^2_\delta$
as a function of height $h$ for several values of $\delta$. The values in the normalization $\sigma^2_\delta$ are taken from the best fit shown as full line in (b). 
Each curve follows a power law with exponent $4/3$, the line corresponding to $h^{4/3}$ is shown as comparison.
(b) The prefactor $\sigma^2_\delta$, where the data are best fits according to (\ref{msd2}). The solid line used for the collapse in (a) follows the prediction $(\delta^2 +\epsilon^2 )^{4/3}$ with fitted $\epsilon =0.40$, which is compatible with the fit in Fig.~\ref{KPZcoefficient}.
}
\end{figure}

\subsection{\label{sec:level4}Domain boundaries}

In this section we derive the superdiffusive behaviour of the domain
boundaries between the species from the scaling properties of the interface.
Since the boundaries grow locally perpendicular to the rough surface, they are expected to be superdiffusive, which has been shown for a solid on solid growth model in \cite{Saito1995} and has been observed in \cite{Hallatschek2007} for experimental data. In order to confirm this quantitatively for our model, we perform simulations with initial conditions $B_1 (0)=\{ 1,\ldots ,[L/2]\}$ and $B_2 (0)=\{ [L/2]+1,\ldots,L\}$, i.e.\ the initial types are all red on the left half and all green on the right half of the linear system. Therefore we have two sector boundaries $X_1$ and $X_2$ with initial positions $X_1(0) =1/2$ and $X_2(0) =[L/2]+1/2$. After growing the whole cluster, we define the boundary as a function of height via the leftmost particle in a strip of width 2 and medium height $h$:
\begin{eqnarray}\label{sbs}
X_1(h) &=\min\big\{ x_p +1/2 :|y_p -h|<1,\, p\in B_1 \big\}  \nonumber\\
X_2(h) &=\min\big\{ x_p +1/2 :|y_p -h|<1,\, p\in B_2 \big\}\ ,
\end{eqnarray}
where we take the periodic boundary conditions into account. The simulations are performed on a system of size $L=1000$, and run until a time of $t=2000$, which is well before the expected time of annihilation, which
is of order $L^{3/2}$ proportional to the saturation timescale in the KPZ class. Therefore we can treat the sector boundaries as two independent realizations of the boundary process $\big(X(h) :h\geq 0\big)$.

As has been noted already in \cite{Saito1995}, this process is expected to follow the same scaling as the lateral correlation length. For the mean square displacement
\be \label{msd}
    M(h):=\left\langle \big(X(h)-X(0)\big)^2 \right\rangle 
\ee
we therefore get with (\ref{paral}) and (\ref{d2nu}), using the linear relationship between $h$ and $t$,
\be\label{msd2}
M(h)\approx \sigma^2_\delta \, h^{2H} \sim \xi_\parallel^2 (h).
\ee
Here $\sigma^2_\delta \propto (\delta^2 +\epsilon^2 )^{4/3}$ and the Hurst exponent is $H=2/3$, which quantifies the superdiffusive scaling of the mean square displacement (\ref{msd}). This prediction is in very good agreement with data for the scaling of $M(h)$ and its prefactor as presented in Fig.~\ref{boundary}, and the fit value for $\epsilon^2$ is consistent with the one in Fig.~\ref{KPZcoefficient}. As before, for $D/(2\nu )$ the $\delta$-dependence is absorbed by the prefactor, and the power law exponent $4/3$ for $M(h)$ remains unchanged from standard KPZ behaviour studied also in \cite{Saito1995}.

\begin{figure}[ht!]
\begin{center}
\subfigure[]{\label{densitypdf}\includegraphics[clip,width=0.9\columnwidth]{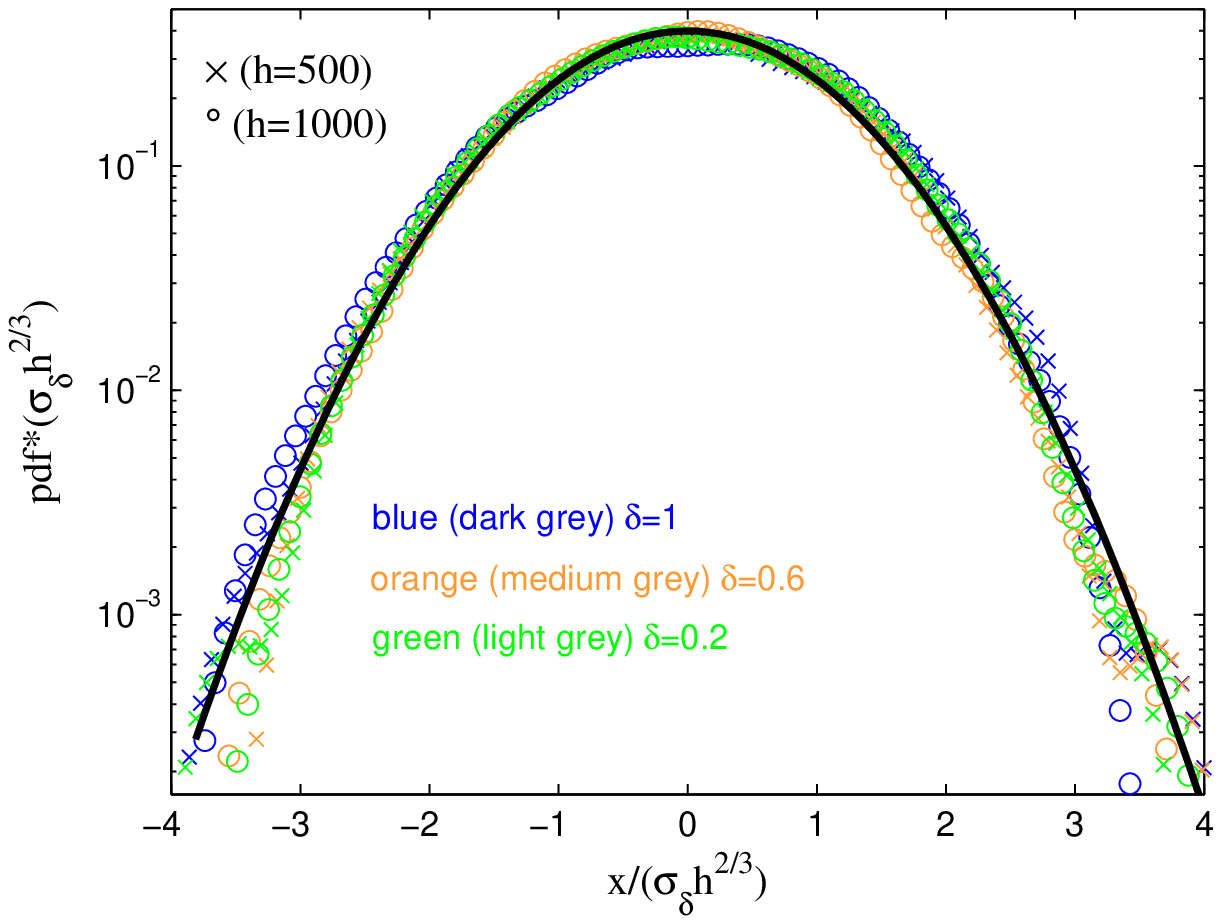}}
\subfigure[]{\label{covfig}\includegraphics[clip,width=0.9\columnwidth]{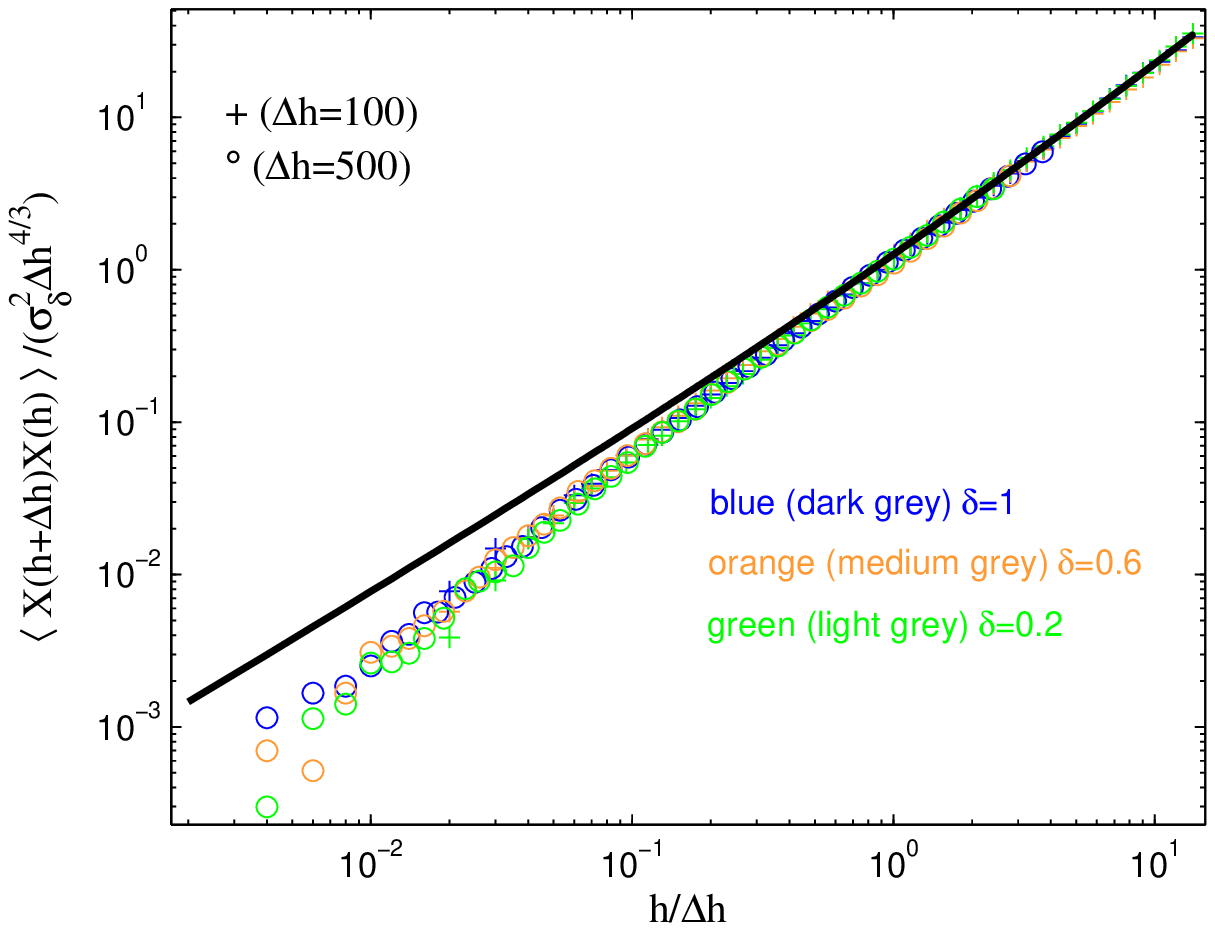}}
\end{center}
\caption{\label{densityfig}
(Color online) The sector boundary $X(h)$ behaves like a fractional Brownian motion. 
(a) The standardized probability density function (pdf) of $X(h)$ as a function of the rescaled argument $x/(\sigma_\delta h^{2/3})$ for different heights $h$ and values of $\delta$.
The black solid parabola is the
pdf of a standard Gaussian on a logarithmic scale. (b) The covariance function $\big\langle X(h+\Delta h)X(h)\big\rangle$ shows the behaviour (\ref{covx}), which is plotted as the solid black curve. After rescaling we get a data collapse as a function of $h/\Delta h$, which agrees well with the prediction if $h$ is sufficiently large and the flat boundary conditions become irrelevant. Data are averages over $1000$ realizations and the error bars are comparable to the size of the symbols.
}
\end{figure}

We can further investigate the law of the process $\big(X(h):
h\geq 0\big)$. The data presented in Fig.~\ref{densityfig}(a) clearly support that
$X(h)$ is a Gaussian process. A fractional Brownian motion with stationary increments seems to be a natural model for the $X(h)$ in the KPZ scaling window. This is confirmed by the behaviour of the correlation function $ \langle X(h+\Delta h)X(h) \rangle$, which is shown in Fig.~\ref{densityfig}(b) for various $\delta$ and two values of the lag $\Delta h>0$. For a fractional Brownian motion with mean square displacement (\ref{msd2})
we expect
\begin{equation} \label{covx}
\big\langle X({h{+}\Delta h})X(h) \big\rangle \approx \frac{\sigma_\delta^2}{2}\big( (h{+}\Delta h)^{2H} {+} h^{2H} {-} |\Delta h|^{2H}\big)
\end{equation}
for all $\Delta h>0$ and $h>0$ sufficiently large to have no effects from the flat initial condition. For simplicity we have assumed here that $X(0)=0$.

This is in good agreement with the data, and we conclude that the sector boundaries can be modeled by fractional Brownian motions with superdiffusive Hurst exponent $H=2/3$ and a $\delta$-dependent prefactor $\sigma_\delta$ (\ref{msd2}). We note that the exponent $H=2/3$ has also been observed in experiments \cite{Hallatschek2007}.

\subsection{\label{sec:level5}Sector patterns}

In \cite{Grosskinsky2010}, and also \cite{Hallatschek2010,Korolev2010} under the assumption of diffusive scaling, it was shown how the understanding of the single
boundary dynamics leads to a prediction for sector statistics for well-mixed
initial conditions. In this section we shortly review this approach and show that it
carries over straight away to systems with $\delta <1$. The sector boundaries $X_i (h)$ interact by diffusion limited annihilation which drives a coarsening process, as can be seen in Fig.~\ref{fig:linear} for two linear populations with different values of
$\delta$. Both systems have the same initial condition with a flat line of particles of independently chosen types, and the finer coarsening patterns for smaller values of $\delta$ result from the reduced boundary fluctuations due to the prefactor $\sigma_\delta$ (\ref{msd2}).
 
\begin{figure}[t]
\begin{center}
\includegraphics[width=0.9\columnwidth]{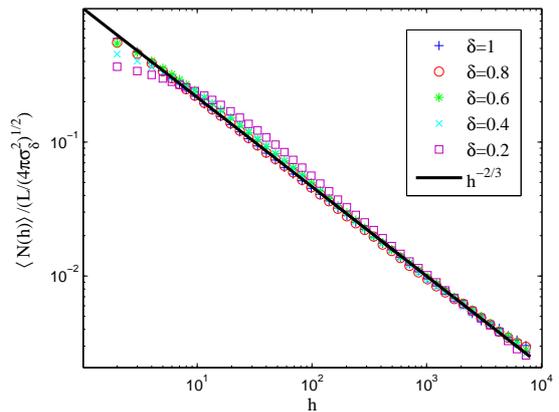}
\end{center}
\caption{\label{sectorsfig}
(Color online) The average number of sector boundaries $\langle N(h) \rangle$ follow a power law (\ref{nofh}) with exponent $-2/3$, which is indicated by the full line. The data are plotted for a system size $L=1500$ and various values of $\delta$ (see legend), and collapse on the function $h^{-2/3}$ when rescaled by $L/\sqrt{4\pi \sigma_\delta^{2}} $. Data are averages over $500$ realizations and the error bars are comparable to the size of the symbols.
}
\end{figure}

Let $N(h)$ be the number of sector boundaries at height $h\geq 0$ as defined in (\ref{sbs}).
For systems of diffusion limited
annihilation \cite{Sasaki2000,Alemany1995} it is known that $N(h)$ is inversely proportional to the root mean square displacement, and decays according to
\be\label{nofh}
  \langle N(h) \rangle \approx
  \frac{1}{\sqrt{4\pi M(h) }} \sim \frac{1}{\sigma_\delta}\, h^{-2/3} \ .
\ee
This prediction is confirmed in Fig.~\ref{sectorsfig}, where we plot $\langle
N (h)\rangle$ for various $\delta$, and obtain a data collapse by multiplying the data with $\sqrt{4\pi\sigma_{\delta}^{2}} /L$, \cite{Alemany1995}. We include the system size $L$ in the rescaling so that rescaled quantities are of order $1$, and all data collapse on the function $h^{-2/3}$ without prefactor.
 
Using (\ref{nofh}), we can predict the expected number of sector boundaries at the final height in the simulations shown in Figure~\ref{fig:linear}. For $\delta =1$, the final height is $h \approx 70$ leading to $\langle N(h) \rangle\approx 7.6$, and for $\delta =0.1$, $h \approx 40$ with $\langle N(h) \rangle\approx 32$. These numbers are compatible with the simulation samples shown which have $6$ and $34$ sector boundaries remaining, respectively.

In general, diffusion limited annihilation is very well understood, and there are exact formulas also for higher order correlation functions \cite{munasingheetal2006}, which can be derived from the behaviour of a single boundary (\ref{msd2}). This demonstrates that the behaviour of populations is fundamentally the
same for all values of $\delta$ and characterized by the KPZ universality class, and the observed difference in coarsening patterns can be explained by the functional behaviour of the prefactors.

\section{\label{sec:bio}Realistic reproduction times}

In this section we study the effect of more realistic reproduction time distributions on the sectoring patterns, and how they can be effectively described by the previous $\delta$-dependent family of distributions in terms of their variation coefficient. As an example, we focus on \textit{S.~cerevisiae}, which is one of the species included in \cite{Hallatschek2007}, and for which reproduction time statistics is available \cite{Cole2004,Cole2007,Karen2008} by the use of time lapsed microscopy. \textit{S.~cerevisiae} cells have largely isotropic shape so that spatial correlations during growth should be minimal, fitting the assumptions of our previous model.
However, the results of this section cannot be applied directly to quantitatively predict the patterns in Fig.~\ref{fig:fluorescent}, since the variation coefficients under the experimental conditions in \cite{Hallatschek2007} are not known to us.

When yeast cells divide, the mother cell forms a bud on its surface which separates from the mother after growth to
become a daughter cell. The mother can then immediately restart this reproduction
process, whereas the daughter cell has to grow to a certain size in order to
be classed as a mother and to be able to reproduce. We denote this time to maturity by
$T_m$ and the reproduction time of (mother) cells by $T_r$.

\begin{figure}[t]
\begin{center}
\subfigure[]{\label{realisticb1}\includegraphics[width=0.8\columnwidth]{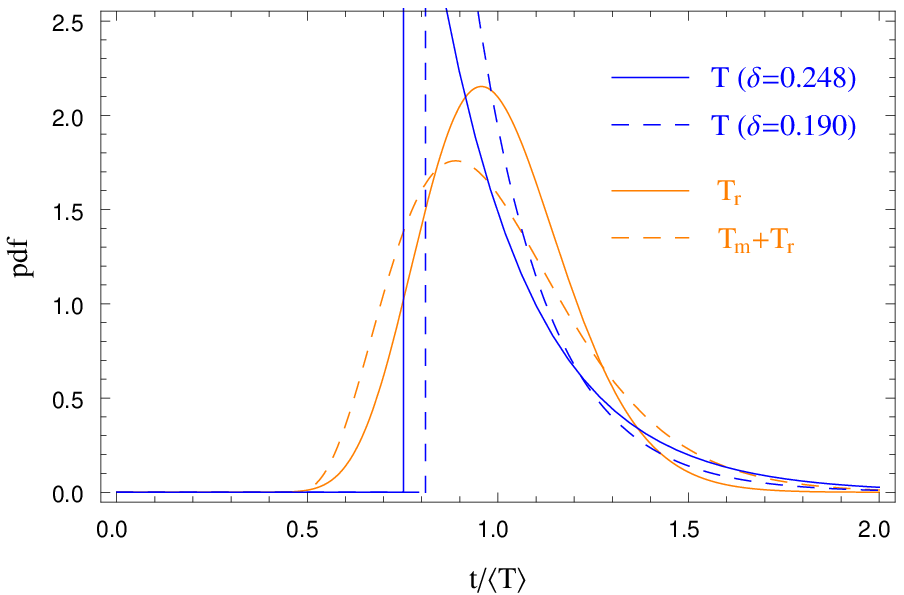}}
\subfigure[]{\label{realisticb2}\includegraphics[clip,width=0.9\columnwidth]{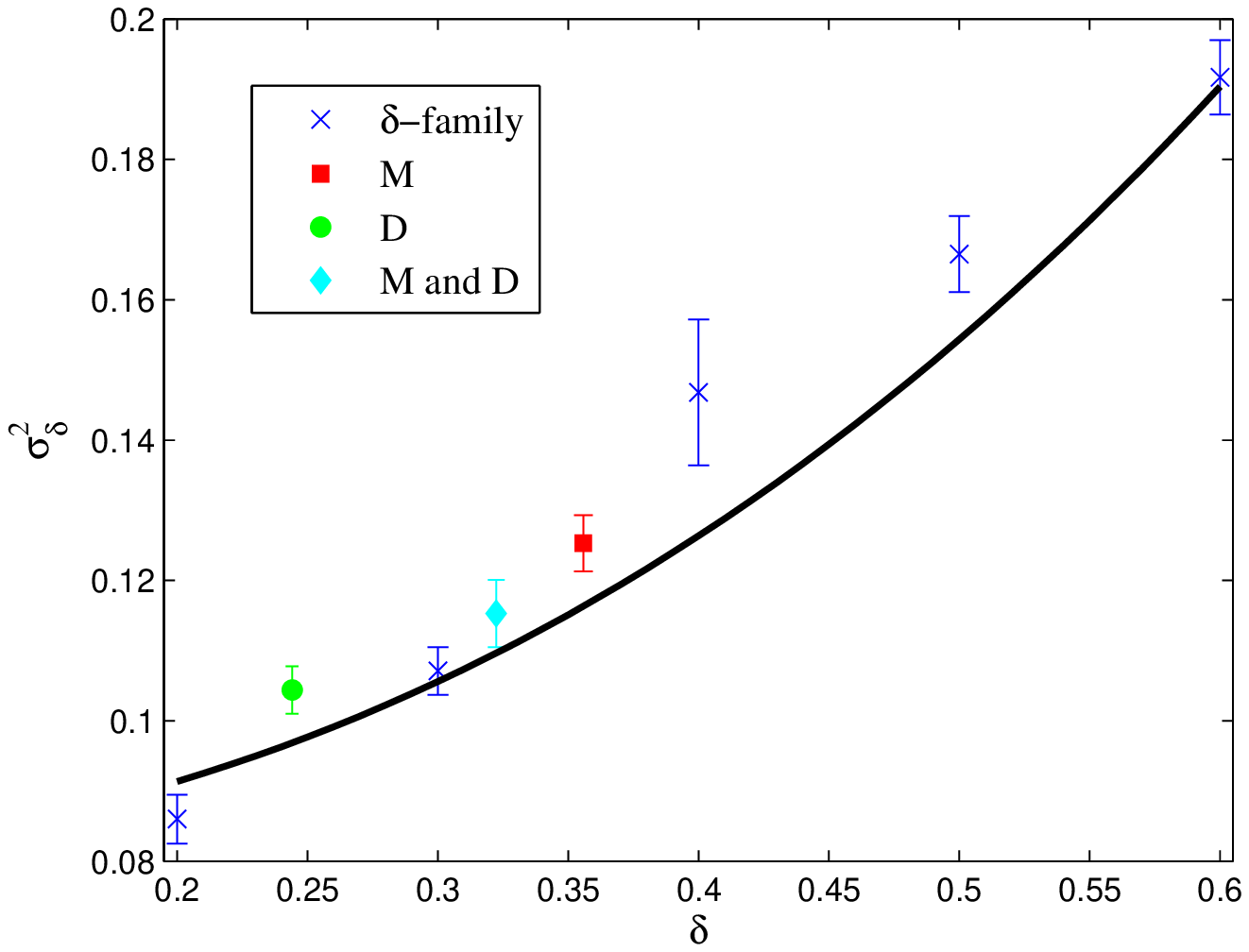}}

\end{center}
\caption{\label{realistic_b}
(Color online) Comparison of realistic reproduction times with the $\delta$ model. (a) The probability density functions of reproduction times of mother cells $T_r$ (full orange line/light grey) and daughter cells $T_m +T_r$ (dashed orange line/light grey) with normalized mean compared to $T$ from (\ref{repro}) with corresponding $\delta$ (blue/dark grey). (b) 
The prefactor of the mean square displacement $\sigma_\delta^{2}$ as introduced in (\ref{msd2}) and Fig.~\ref{boundary}.
The data correspond to reproduction times $T_r$ for all cells (denoted M), $T_m +T_r$ for all cells (denoted D) and the most realistic mixed model (denoted M and D) as explained in the text.
All these cases are consistent with previous results from Fig.~\ref{boundary}.
}
\end{figure}

The results in \cite{Karen2008,Cole2007,Cole2004,Powell1956} suggest that Gamma distributions are a reasonable fit for the statistics for $T_m$ and $T_r$, where
\begin{equation}
  T_r \quad\text{is distributed as}\quad\rho_{0} +\text{Gamma} (\rho_{1}, \rho_{2}) \label{Mother},
\end{equation}
with delay $\rho_0 >0$. The parameters $\rho_1 ,\rho_2$ denote the shape and scale of the Gamma distribution, which has a probability density function
$$
   f_{\rho_1 ,\rho_2} (t)=t^{\rho_{1}-1}\frac{\exp{\left(-t/\rho_{2}\right)}}{\Gamma(\rho_{1})
    \rho_{2}^{\rho_{1}}}\ ,\quad t\geq 0\ .
$$
The time to maturity is
\begin{equation}
    T_m \quad\text{distributed as} \quad \text{Gamma} (\rho_{3}, \rho_{4}) \label{maturity},
\end{equation}
and in \cite{Karen2008} data are presented for which the parameters can be fitted to $\rho_{0}\approx 1.0$, $\rho_{1}\approx1.7$, $\rho_2\approx0.51$, $\rho_{3}\approx 9$ and $\rho_4 \approx 0.3$. The unit of $\rho_{0}$, $\rho_{2}$ and $\rho_{4}$ are hours and $\rho_{1}$, $\rho_{3}$ are dimensionless numbers.\\ 
The random variables $T_m$ and $T_{r}$ may be assumed to be independent and the time until a newly born daughter cell can reproduce is distributed as the sum $T_m +T_r$. Note that the expected value of reproduction times $\langle T_r \rangle =\rho_0 +\rho_1 \rho_2 =1.86$ is smaller than that for times to maturity $\langle T_m \rangle =\rho_3 \rho_4 =2.52$ but of the same order. The real time scale for these numbers is hours, but we are only interested in the shape of the distributions rescaled to mean $1$ like our previous model.

The distribution (\ref{repro}) of $\delta$-dependent reproduction times can be written as $T = 1-\delta +\text{Gamma}(1,\delta)$, since exponentials are a particular case of Gamma random variables with shape parameter $1$. The reproduction time $T_r$ of mother and $T_m +T_r$ of daughter cells are also unimodal with a delay, and very similar in shape to $T$ in our model. This can be seen in Fig.~\ref{realistic_b}(a), where we plot the probability densities renormalized to mean $1$. Analogous to (\ref{delta}), we can compute the variation coefficients of $T_r$ and $(T_m +T_r)$, which turn out to be $0.356$ and $0.244$, respectively.
To confirm that the behaviour of sector boundaries can be well predicted by the variation coefficient, we present data of three simulations in Fig.~\ref{realistic_b}(b): one with reproduction times $T_r$ for mother and $T_d +T_r$ for daughter cells as explained above, one with $T_r$ for all cells, and one with $T_r +T_m$ for all cells. The mean square displacement $M(h)$ for these models also shows a power law with exponent $4/3$ analogous to Fig.~\ref{boundary}, and the prefactors $\sigma_\delta$ match well with our simplified model.

To estimate the variation coefficient in the model with mother and daughter cells, we measure the fraction of reproduction events of daughter cells to be $p_d =0.88$, and $p_m =0.12$ for mother cells. The reproduction time of the union of mother and daughter cells is then taken as
\be\label{map}
T \quad\text{distributed as} \quad \Theta (T_m +T_r ) + (1-\Theta ) T_r \ ,
\ee
where the independent Bernoulli variable $\Theta=\text{Be}(p_d )\in\{ 0,1\}$ indicates reproduction of a daughter. The variation coefficient of $T$ turns out to be $0.322$.
In all three combinations of realistic reproduction times we find that the generic
family of $F_{\delta}$ introduced in (\ref{repro}) provides a good approximation for the properties of domain boundaries in simulations. We expect this method of mapping realistic reproduction time distributions to our $\delta$-dependent family to hold for a large class of microbial species which have similar unimodal distributions.

\section{\label{sec:level7}Conclusion}

We have introduced a generalization of the Eden growth model with competing species, regarding the reproduction time statistics of the individuals. This is highly relevant in biological growth phenomena, and can have significant influence on the sectoring patterns observed e.g.\ in microbial colonies. Although growth of immotile microbial species is the prime example, our results also apply to more general phenomena of space limited growth with inheritance, where the entities have a complex internal structure that leads to non-exponential reproduction times, such as colonization/range expansions or epidemic spreading of different virus strands. Our main result is that, as long as the reproduction time statistics have finite variation coefficients, the induced correlations are local and the macroscopic behaviour of the system is well described by the KPZ universality class. The dependence of the relevant parameters in that macroscopic description on the variation coefficient (a microscopic property of the system) is well understood by simple heuristic arguments, which we support with detailed numerical evidence. 

Figures~\ref{fig:circular} and \ref{fig:linear} illustrate that changes in the variation coefficient $\delta$ of reproduction times lead to significant changes in the competition growth patterns in our model, and we are able to quantitatively predict this dependence. We studied the effects of reproduction time statistics in a generalized Eden model, isolated from other influences such as shape of the organisms or correlations between mother and daughter cells which might be relevant in real applications. In that sense our results are of a theoretical nature. However, they indicate that the variation coefficient of reproduction times can have a strong influence on observed competition patterns. This coefficient has been measured for various species in the literature, where it is found that it depends on experimental conditions such as type of strain, concentration of nutrients, temperature etc.\ \cite{Elfwing2004,Rahn1932,Slater1977}. For example, it was found that for \textit{S.~cerevisiae} the coefficient for mother cells can vary between $\delta\approx 0.12 - 0.38$ and for daughter cells $\delta\approx 0.19 - 0.28$ depending on concentrations of guanidine hydrochloride. It has also been observed that $\delta$ can be as small as $0.047$ for these yeast type organisms \cite{Gaskins2008}. For \textit{E.~coli} values of $\delta\approx 0.32 - 0.51$ have been observed in \cite{Niven2006,Elfwing2004,Wakamoto2005}, which is larger, and compatible with the observations in Fig.~\ref{fig:fluorescent}. But for the experimental conditions in \cite{Hallatschek2007} with pattern growth the coefficient has not been determined and therefore the results in this paper cannot be readily applied to explain the differences in competition patterns between \textit{S.~cerevisiae} and \textit{E.~coli}. In particular, the latter have anisotropic rod shape which has probably a strong influence and is currently under investigation in the group of O.\ Hallatschek. 
Another rod shaped bacterium, \textit{Pseudomonas aeruginosa}, has variation coefficient $\delta \approx 0.14-0.2$ \cite{Powell1956,Niven2006}. This bacterium along with \textit{E.~coli} belongs to the gram-negative bacteria family. Despite obvious similarities between \textit{P.~aeruginosa} and \textit{E.~coli} in the shape of their cells, their colonies display morphological differences \cite{Korolev2011}, which fit qualitatively into our results, and would be another interesting example for a quantitative study.

In general, it is an interesting question if the simple mechanism of time correlations due to reproduction time statistics with variable variation coefficients is sufficient to quantitatively explain sectoring patterns in real experiments. We are currently investigating this for \textit{S.~cerevisiae} which are approximately of isotropic shape, in order to quantify the influence of other factors on growth patterns, such as correlations between mother and daughter cells. For example, an effective attraction between cells which is often observed in the growth of microbial colonies would influence the growth directions, and further smoothen the surface and the fluctuations of sector boundaries. For future research, it should also be possible to describe spatial effects due to non-isotropic particle shapes with the methods used in this paper.

\section*{Acknowledgments}

The authors thank O.~Hallatschek for useful discussions. This work was supported by the Engineering and Physical Sciences Research Council (EPSRC), grant no.\ EP/E501311/1.

\appendix

\section{\label{epsilon} Effect of geometric disorder}

The squared variation coefficient $\epsilon^2$ in (\ref{xiperp}) due to geometric disorder has been consistently fitted to values around $0.4$ with our data in Section \ref{sec:results}. This value is compatible with the following very simple argument. Consider a single growth event around an isolated spherical particle with diameter $1$, with direction $\alpha$ chosen uniformly in a cone with opening angle $\pi /2$ around the vertical axis. This leads to $\langle\Delta y_i \rangle =\int_{-\pi /2}^{\pi /2} \cos\alpha \frac{d\alpha}{\pi}\approx 0.64$ and
\be
\epsilon^2 \approx \Big(\int_{-\pi /2}^{\pi /2} \cos^2\alpha\frac{d\alpha}{\pi} -\langle\Delta y_i \rangle^2 \Big)\Big/ \langle\Delta y_i \rangle^2 \approx 0.23\ ,
\ee
which is of the same order as the fitted values. Choosing only a slightly larger opening angle $0.55\pi$ of the cone leads to $\epsilon^2 \approx 0.39$ and $\langle\Delta y_i \rangle\approx 0.57$. These are in good agreement with the fitted values and with measurements of $\langle\Delta y_i \rangle$ (not shown). The latter show some dependence on $\delta$, related also to the compactness of growth as seen in Figs.~\ref{fig:circular} and \ref{fig:linear}, but this does not contribute to our results on a significant level so we ignore this dependence. Actual growth events in the simulation are of course often obstructed by neighbouring particles, but the right order of magnitude of the parameters can be explained by the basic argument above.

\section{\label{derivecorrelation} Deriving the correlation function $C(l,t)$}

We use the mode coupling method \cite{Amar1992}, in order to find an exact analytical expression of the  correlation function Eq.~(\ref{correlationfunction}) as shown in Eq.~(\ref{correlationresult}). The idea of the mode coupling approximation is that properties of solutions of the KPZ equation (\ref{kpz}) may be derived by first considering the linear Edwards-Wilkinson equation \cite{Edwards1982} for $\lambda =0$. We further consider the co-moving frame, so that $v_{0}=0$, and the equation then reads
\be
  \partial_{t} y(x,t)= \nu \Delta y(x,t)
  + \sqrt{D}\eta(x,t)
  \label{EW} .
\ee
We denote by
$$\hat{y}(k,t)=\int_{-\infty}^{\infty} dx\, y(x,t) e^{-ikx}$$
the Fourier transform of the function $y(x,t)$. 
The evolution of the function $\hat{y}(k,t)$ satisfies \be
  \partial_{t} \hat{y}(k,t)= -\nu k^{2} \hat{y}(k,t)
  + \sqrt{D}\,\hat{\eta}(k,t)
  \label{FEW}.
\ee
Here $\hat{\eta}(k,t)$ is the spatial Fourier transform of the white noise $\eta(x,t)$, where $\hat{\eta}(k,t)$ has a mean $0$ with correlations
\be
    \big\langle \hat{\eta}(k,t)\hat{\eta}(k',t') \big\rangle =\frac{1}{2\pi}\delta(k+k')\delta(t-t').
\ee
A formal solution of (\ref{FEW}) can be obtained, and after inverse Fourier transform this leads to

\be
y(x,t)= \sqrt{D} \int_{-\infty}^{\infty}dk\, e^{ikx} \int_{0}^{t} ds\,\hat{\eta}(k,s) e^{-\nu k^{2}(t-s)}\ . \label{EWsol}
\ee 
The correlation function $C(l,t)$ defined in (\ref{correlationfunction}) can be represented as
\be
C(l,t)^{2}= 2\int_0^L dx\big\langle y(x,t)^2{-}y(l{+}x,t)y(x,t)\big\rangle\ . \label{correlationfunction1}
\ee
Using the solution (\ref{EWsol}) we can compute
\be
\int_{0}^{L}\!\!\!\!\! dx\big\langle y(l{+}x,t)y(x,t)\big\rangle = \frac{D}{2 \nu \pi}\int_{0}^{\infty}\!\!\!\!\! dk\, \frac{\cos(kl)}{k^{2}}[1{-} e^{{-}2\nu k^{2}t}]\ ,\label{hh}
\ee
where we have used that the Fourier transform is even in $k$. Taken together, this leads to an expression for the correlation function (\ref{correlationfunction}) of the Edwards-Wilkinson equation
\be
C(l,t)^{2} =\frac{D}{\nu \pi}\int_{0}^{\infty}\!\!\! dk\, k^{-2}\big( 1{-}\cos(kl)\big) [1{-} e^{{-}2\nu k^{2}t}]\ \label{covEW}.
\ee 

In order to compute the correlation function for the KPZ equation (\ref{kpz}) we substitute length scale dependent parameters $D(k)$ and $\nu(k)$ into (\ref{covEW}), which are obtained from the renormalization group flow equations \cite{Kardar1986,Amar1992,Barabasi1995}. In $d=2$ dimensions these are given by
\begin{eqnarray}\label{reno}
\nu(k)&=&\nu_{1} [(1-\alpha_{B})+\alpha_{B}/k]^{1/2}\ ,\nonumber\\
D(k)&=&D_{1}[(1-\alpha_{B})+\alpha_{B} /k]^{1/2}\ ,
\end{eqnarray}
and $\lambda(k)=\lambda_{1}$, where
$$\alpha_{B}=\frac{\lambda_{1}^{2}D_{1}}{4\pi^{2} \nu_{1}^{3}}\ .$$ 
Here $(\lambda_{1}, \nu_{1},D_{1})$ are the parameters for $k=1$ where no renormalization has taken place. Plugging this into (\ref{covEW}) and only considering the most dominant terms, we obtain
\be
C(l,t)^{2} =\frac{D_{1}}{\nu_{1} \pi}\int_{0}^{\infty}\!\!\! dk\, k^{-2}\big( 1{-}\cos(kl)\big) [1{-} e^{{-}B k^{3/2}t}]\ ,  \label{covkpz} \ee
where $B=2\nu_{1}\alpha_{B}^{1/2}=\frac{\sqrt{2}}{\pi}\lambda \sqrt{D_{1}/2\nu_{1}}$\ .\\
If we take $t\rightarrow \infty$ in Eq.~(\ref{covkpz}) we get
\begin{eqnarray}
C(l,t)^{2} \rightarrow \frac{D_{1}}{\nu_{1} \pi}\int_{0}^{\infty}\!\!\! dk\, k^{-2}\big( 1-\cos(kl)\big) =
\frac{D_{1}}{2\nu_{1}} \, l\ .
\end{eqnarray}
With (\ref{reno}) $D/\nu= D_{1}/\nu_{1}$ is independent of the scale $k$, and thus
\be\label{exactpre}
C(l,t)\approx \Big(\frac{D}{2\nu}\, l\Big)^{1/2} \quad\mbox{for}\quad l\ll \xi_{\parallel}(t)\ .
\ee
For finite time, numerical integration of (\ref{covkpz}) in the large $l$ limit gives
$$\lim_{l\rightarrow \infty} C(l,t)^{2} \approx 2.7\, \frac{D}{\nu \pi}\, (Bt)^{2/3}. $$ 
Together with (\ref{exactpre}) and the definition (\ref{correlationfunction}) of the correlation length this leads to
\be
\lim_{l\rightarrow \infty} C(l,t) \approx \Big( 5.4 \times 2^{1/3} \Big(\frac{D}{2\nu}\Big)^{4/3} \pi^{-5/3} (\lambda t)^{2/3}\Big)^{1/2}\ , \label{covkpzlimit}
\ee
and
\be
\xi_{\parallel}(t) \approx 5.4 \times 2^{1/3} \Big(\frac{D}{2\nu}\Big)^{1/3} \pi^{-5/3} (\lambda t)^{2/3}\ .\label{xiparallel}
\ee

\end{document}